\documentclass[ 
    aps,
    onecolumn,
    groupedaddress,
    nofootinbib,
    longbibliography,
    balancelastpage,
    notitlepage,
    superscriptaddress
]{revtex4-2}

\usepackage[
    colorlinks=true,
    urlcolor=blue,
    anchorcolor=blue,
    citecolor=blue,
    filecolor=blue,
    linkcolor=blue,
]{hyperref}

\usepackage[capitalise]{cleveref}
\usepackage{verbatim}
\usepackage{siunitx}
\DeclareSIUnit{\parsec}{pc}

\usepackage{aas_macros}
\newcommand{\refcite}[1]{Ref.~\cite{#1}}
\newcommand{\refscite}[1]{Refs.~\cite{#1}}

\begin{document}

\title{Black hole remnants are not too fast to be dark matter}

\author{Benjamin V. Lehmann}
\email{benvlehmann@gmail.com}

\author{Stefano Profumo}
\email{profumo@ucsc.edu}

\affiliation{Department of Physics, University of California Santa Cruz, 1156 High St., Santa Cruz, CA 95064, USA}
\affiliation{Santa Cruz Institute for Particle Physics, 1156 High St., Santa Cruz, CA 95064, USA}

\begin{abstract}
    We comment on recent claims that recoil in the final stages of Hawking evaporation gives black hole remnants large velocities, rendering them inviable as a dark matter candidate. We point out that due to cosmic expansion, such large velocities at the final stages of evaporation are not in tension with the cold dark matter paradigm so long as they are attained at sufficiently early times. In particular, the predicted recoil velocities are robustly compatible with observations if the remnants form before the epoch of big bang nucleosynthesis, a requirement which is already imposed by the physics of nucleosynthesis itself.
\end{abstract}

\maketitle

Remnants of black hole evaporation have long been considered a potential dark matter candidate \cite{MacGibbon:1987my,Chen:2002tu}. Recently, however, \refcite{Kovacik:2021qms} argued that such remnants are incompatible with cold dark matter. The reasoning is as follows: such remnants are expected to form if evaporation stalls at a mass $M_{\mathrm R}$ near the Planck scale $M_{\mathrm P}$. In the final stages of evaporation, the black hole will thus have an extremely high temperature, as $T\sim M_{\mathrm P}^2/M_{\mathrm R}\sim M_{\mathrm P}$. Such a black hole sheds an appreciable fraction of its mass with the emission of a small number $N_{\mathrm q}$ of highly energetic quanta, effectively performing a random walk in momentum due to the successive recoils. The typical momentum of the black hole after such a random walk scales as $1/\sqrt{N_{\mathrm q}}$, and with a few well-motivated assumptions regarding the final stages of evaporation, \refcite{Kovacik:2021qms} finds that typical remnants acquire a velocity $v\sim0.1$ as a result of the process. \refcite{Kovacik:2021qms} states that these velocities are in tension with the cold dark matter paradigm. This motivated the more recent analysis of \refcite{Gennaro:2021amf}, which found that such bounds could be evaded in a certain class of modifications to the physics of evaporation.

Here we point that, in fact, the recoil velocity does not constrain typical models of black hole remnants as dark matter: crucially, the last step of the argument neglects the effects of cosmic expansion. These remnants may indeed be produced `warm', with semi-relativistic speeds, but the population will inevitably cool as the Universe expands. The cold dark matter paradigm requires only that they slow to non-relativistic speeds by the time observable structures begin to form, which is much less constraining. Indeed, in standard particle dark matter scenarios, the dark matter species is relativistic at early times, being in thermal equilibrium at temperatures much greater than its mass \cite{Profumo:2017hqp}. Due to cosmic expansion, such thermal relics eventually become cold. Nor does this effect require thermal equilibrium: consider, for instance, the cosmic neutrino background, which thermally decoupled at a temperature of $\mathcal O(\SI{1}{\mega\electronvolt})$, and which today is non-relativistic \cite{Irvine:1983nr}. The momentum of each individual particle, regardless of its nature, is inversely proportional to the cosmic scale factor \cite{Kolb:1990vq}.

A species that is cold today may still have a momentum history incompatible with structure formation, since the smallest dark matter structures in the late universe formed at early times. In general, a dark matter structure of a given size starts to form when the cosmological horizon grows to encompass the length scale of the corresponding density perturbations. If dark matter is relativistic and kinetically decoupled at this time, then the dark matter particles will `free-stream' rather than immediately forming a bound structure. Thus, the smallest dark matter structures observed today correspond to the earliest times at which dark matter is required to be non-relativistic. The situation considered here is analogous to particle dark matter models in which the dark matter species undergoes thermal decoupling while still relativistic (`warm dark matter'), which can indeed suppress the formation of structure on small scales \cite{Bode:2000gq,Berezinsky:2003vn,Green:2003un,Green:2005fa,Bringmann:2009vf}.

The scale at which structure formation is suppressed is fixed by the comoving free streaming length $\ell_{\mathrm{fs}}$, i.e., the comoving length scale over which a typical particle travels after free streaming begins. Inhomogeneities on smaller scales are erased, and this produces an exponential cutoff in the matter power spectrum at a corresponding wavenumber $k_{\mathrm{fs}} \sim 1/\ell_{\mathrm{fs}}$. For particle dark matter that begins free streaming after kinetic decoupling, this cutoff is estimated by \refcite{Green:2005fa} as
\begin{equation}
    \label{eq:kfs}
    k_{\mathrm{fs}} \simeq
    \left(\frac{m_\chi}{T_{\mathrm{kd}}}\right)^{1/2}
    \frac{a_{\mathrm{eq}} / a_{\mathrm{kd}}}
         {\log(4a_{\mathrm{eq}}/a_{\mathrm{kd}})}
    \frac{a_{\mathrm{eq}}}{a_0}
    H_{\mathrm{eq}}
    ,
\end{equation}
where $m_\chi$ is the dark matter mass; $T_{\mathrm{kd}}$ is the dark matter temperature at kinetic decoupling; $a_{\mathrm{eq}}$ and $H_{\mathrm{eq}}$ denote the scale factor and the Hubble parameter at matter-radiation equality; $a_{\mathrm{kd}}$ denotes the scale factor at kinetic decoupling; and $a_0 = 1$ denotes the scale factor today. Here the ratio $m_\chi / T_{\mathrm{kd}}$ appears as a parameter of the phase space distribution of dark matter in thermal equilibrium. While the black hole remnants are not in thermal equilibrium, their phase space distribution is very similar: a set of random walks in momentum space produces momenta that are approximately normally distributed, which corresponds exactly to the Maxwell--Boltzmann distribution in the non-relativistic limit. Thus, for an order of magnitude estimate, we can parametrize the mass-to-temperature ratio by the rms velocity $\bar v$, using $m_\chi/T \approx 3/\bar v^2$. Then, writing $H_{\mathrm{eq}}$ in terms of the present-day matter density $\Omega_{\mathrm M}$, \cref{eq:kfs} becomes
\begin{equation}
    k_{\mathrm{fs}} \simeq
    \frac{\sqrt{6}H_0\Omega_{\mathrm{M}}^{1/2}a_{\mathrm{eq}}^{1/2}}
         {\bar v_ia_i\log(4a_{\mathrm{eq}}/a_i)}
    \approx \SI{0.18}{\mega\parsec^{-1}}\left(\frac{0.1}{\bar v_i}\right)
        \left(\frac{a_{\mathrm{eq}}}{a_i}
              \frac{1}{\log(4a_{\mathrm{eq}}/{a_i})}\right)
    ,
\end{equation}
where now $\bar v_i$ denotes the rms velocity of the black hole remnants at the beginning of free streaming, i.e., immediately after they are accelerated by evaporation.

The smallest structures probed by observations today still correspond to scales $k\lesssim \SI{10}{\mega\parsec^{-1}}$ in the matter power spectrum \cite{Abolfathi:2017vfu,Troxel:2017xyo}, so to determine the latest viable formation time of the remnants, we can conservatively impose $k_{\mathrm{fs}} > \SI{10}{\mega\parsec^{-1}}$. Taking $\bar v_i = 0.1$, this translates to the requirement that $a_i \lesssim \num{7e-7}$. In the most extreme case, when $N_{\mathrm q} = 1$ and the recoil momentum is $p \simeq M_{\mathrm R}$, then $E^2 = p^2 + M_{\mathrm R}^2 \simeq 2 M_{\mathrm R}^2$, so $\gamma \simeq \sqrt2$, or $\bar v_i \simeq 1/\sqrt2$. This, in turn, corresponds to the constraint $a_i\lesssim\num{8e-8}$. A comparable bound can be derived from the effective condition that the dark matter velocity today is less than \num{5e-7}, as stated by \refscite{Fujita:2014hha,Morrison:2018xla} based on the results of \refcite{Viel:2005qj}. The momentum $p_0$ of a remnant today is related to the momentum $p_i$ after evaporation by $p_0 = a_ip_i$, so we find that $a_i\lesssim\num{5e-7}$ for $\bar v_i = 1/\sqrt2$.\footnote{This is a numerical coincidence, not a typographical error.}

These upper bounds on $a_i$ might indeed be viewed as constraints on the evaporation history of black hole remnants due to the recoil velocity: in order to account for dark matter, such objects must form at scale factors $a\lesssim\num{e-7}$. However, this is much larger than the scale factor at the epoch of big bang nucleosynthesis (BBN), $a_{\mathrm{BBN}} \simeq \num{2.5e-10}$. BBN is highly sensitive to injected energy, so there are already stringent constraints on active evaporation during this epoch \cite{1978SvA....22..138V,Kohri:1999ex,Carr:2009jm,Acharya:2020jbv}. Typically, in phenomenologically viable models of black hole remnants, evaporation must halt when $a < a_{\mathrm{BBN}}$ to evade these bounds. Therefore, while the recoil velocity can in principle lead to constraints on the evaporation history of black hole remnants as dark matter, the actual constraints derived from this method are subdominant to existing BBN limits.

We conclude that in any model of black hole remnants as dark matter that is observationally consistent with light element abundances, the remnants are automatically `cold' as required by the cold dark matter paradigm. In particular, black hole remnants remain viable as a dark matter candidate.

\section*{Acknowledgments}
We gratefully acknowledge valuable discussions with Samuel Kov\'a\v{c}ik and Yen Chin Ong. BVL and SP are supported in part by DOE grant DE-SC0010107.

\bibliography{main}

\end{document}